\documentclass[reprint]{revtex4-2}
\usepackage{xcolor}
\usepackage{ragged2e}
\usepackage{graphicx}

\newcommand{\deltaSAW}{\delta_\mathrm{SAW}}
\newcommand{\deltaT}{\langle d\rangle_\mathrm{T}}

\begin{document}

\title{High-resolution acoustic field mapping of GHz phononic crystals with atomic force microscopy}

\author{Alessandro Pitanti*}
\affiliation{Paul-Drude-Institut für Festkörperelektronik, Leibniz-Institut im Forschungsverbund Berlin e. V., 5-7, Hausvogteiplatz, Berlin, 10117, Germany}
\altaffiliation{NEST, CNR Istituto Nanoscienze and Scuola Normale Superiore, piazza San Silvestro 12, 56127, Pisa, Italy}
\email{alessandro.pitanti@nano.cnr.it}
\author{Mingyun Yuan}
\affiliation{Paul-Drude-Institut für Festkörperelektronik, Leibniz-Institut im Forschungsverbund Berlin e. V., 5-7, Hausvogteiplatz, Berlin, 10117, Germany}
\author{Simone Zanotto}
\affiliation{NEST, CNR Istituto Nanoscienze and Scuola Normale Superiore, piazza San Silvestro 12, 56127, Pisa, Italy}
\author{Paulo Ventura Santos}
\affiliation{Paul-Drude-Institut für Festkörperelektronik, Leibniz-Institut im Forschungsverbund Berlin e. V., 5-7, Hausvogteiplatz, Berlin, 10117, Germany}


\begin{abstract}
On-chip technology based on acoustic waves is a strong asset in modern telecommunication with the prospects of becoming a cornerstone of next-generation devices. In this context, mapping and manipulating acoustic waves through coherent scattering is pivotal for a non-trivial control of the flow of acoustic energy, which could consequently enable advanced information manipulation. To this end, here a technique for mapping acoustic fields is introduced and used for characterizing $\mu$m-sized phononic crystals defined in GaAs slabs and excited by GHz surface acoustic waves (SAWs) based on atomic force microscopy. It is shown that incoherent scattering excites a wide distribution of modes, which enables the mapping of the dispersion relation of the two-dimensional structures, while the phononic crystal symmetry directly correlates with coherent scattering effects. Enabling the use of acoustic atomic force microscopy and understanding the role of scattering are of paramount importance for  the versatile use of GHz acoustic waves in technological applications, setting the baseline for advanced operations like hyperspectral filtering, beam steering or spatial-division multiplexing.  
\end{abstract}

\maketitle

\section{Introduction}

The last decades have witnessed a rich activity towards the integration of acoustic technologies within electrical circuits in high-frequency hybrid devices. The main role in this trend has been played by surface acoustic waves (SAWs)~\cite{Rayleigh1885}, which are wave propagating along a surface that can be easily introduced in several material platforms exploiting the piezoelectric effect~\cite{RezaAli2020}. Given their high operation frequency (typically up to $\sim$10 GHz) and quality factors, simple, unidimensional SAW delay-line resonators have found wide application as sensors \cite{Devkota2017, Lange2008}, spectral filters \cite{Morgan2010}, and oscillators \cite{Bernardo2002} for telecommunication applications \cite{Hashimoto2000}. Further technological advances have made SAWs a flexible tool for both applied and fundamental physics, the latter reaching, for example, the quantum regime and enabling single particle manipulation \cite{Delsing2019}. In this wide scenario, SAW-based technologies offer novel tools for manipulating information-carrying waves on-chip. Increasing the degree of freedom in the control of SAWs can thus be a relevant key for the transition to 6G technologies: one can imagine, along with an increase in operating frequency, an enhancement of the data transmission rate through wave manipulation with complex spatial or frequency multiplexing, allowing simultaneous on-chip operations, such as filtering, on several communication channels. 

As for electromagnetic waves, acoustic waves can be manipulated by using artificially defined structures in the form of phononic  crystals ($PhC$) or acoustic metasurfaces. The impressive results obtained in advanced light control are presently being translated to acoustic waves with $\mu$m-sized  wavelengths ($\sim$GHz frequencies), propagating on the millimeter to centimeter scale. Interesting proof-of-concepts experiments have already been demonstrated or theoretically proposed, including acoustic holography \cite{Xu2022}, negative acoustic refraction~\cite{He2018} as well as  several topological effects \cite{Cha2018,Zhang2018,Luo2021}, albeit most in the kHz to MHz frequency range, with only a  few exceptions reaching the 100s MHz \cite{Shao2020,Gao2023} and GHz range \cite{Zanotto2022}. The latter frequency ranges are the most promising for integration with modern communication technologies, since they offer mm to cm footprint devices operating at the frequency of the standard electromagnetic carrier waves for wireless telecommunication. Moreover, the acoustic wavelengths of the mostly used semiconductor materials in the GHz range is comparable to those of 3$^{rd}$ telecom window photons. SAW-based structures thus offera powerful platform for photon manipulation \cite{deLima2005} and optomechanical hybrid systems for light-based telecommunication and favouring interesting features such as coherent wavelength conversion \cite{Pitanti2020,Forsch2020,Navarro2022}.
The development of high-frequency phononic crystals requires efficient techniques for probing GHz acoustic field distributions in the sub-$\mu$m length scale. Most of these acoustic devices are routinely measured using either integrated probes \cite{Khelif2003,Fang2016} or techniques based on light interferometry \cite{Kokkonnen2007,Profunser2009,Achaoui2011,Kurosu2018}. While presenting several advantages, the former case does not allow for direct mapping of the vibrational field, while the latter presents intrinsic limitations associated with the finite spatial resolution imposed by the light diffraction limit, as well as  lack of sensitivity to in-plane fields. A powerfull tool for mapping vibrational fields, which so far has only be scarcely used, is the Atomic Acoustic Force Microscopy (AAFM),  a variation of the conventional Atomic Force Microscopy (AFM). This fast-scanning technique was initially introduced for field mapping of kHz to MHz elastic wave structures \cite{Rabe1994} and  subsequently extended to GHz SAWs~\cite{Chilla1997,Kubat2004,Hesjedal2010,Hu2011,Hellemann2022}. The sub-nm lateral sensitivity of AAFM makes it a natural candidate for  field mapping of complex phononic devices operating at  sub-$\mu$m acoustic wavelengths and frequencies in the 10s to 100s GHz range. To the best of our knowledge, however,  AAFM has so far never been applied for investigating GHz modes in $PhC$ devices. 

In this manuscript, we introduce AAFM as a powerful high-resolution tool for mapping sub-$\mu$m fields in GHz $PhC$. In particular, we  report on an AAFM investigation of $PhC$ with different symmetries operating at around 1 GHz. We show that AAFM  allows for recording with a fine level of details the acoustic field distribution within the $\mu$m-sized $PhC$, which are already among the smallest features that can be resolved using visible-light based interferometry. Using the mapping features of AAFM, we analyze isofrequency countours in the reciprocal space, estimating and differentiating between acoustic scattering channels induced by the $PhC$ structure from those originating from pervasive imperfections, the former being ultimately responsible for deterministic wave manipulation. 

The manuscript is organized as follows. After a brief introduction to the device fabrication and theoretical design, we introduce our characterization bench based on AFM. An initial investigation of the SAW on the top surface of the semiconductor substrate is followed by the evaluation and characterization of the mechanical Bloch modes within the suspended phononic crystal membranes.  

\section{Device design and experimental setup}

\begin{figure}[h!]
\centering
\includegraphics[width=1\linewidth]{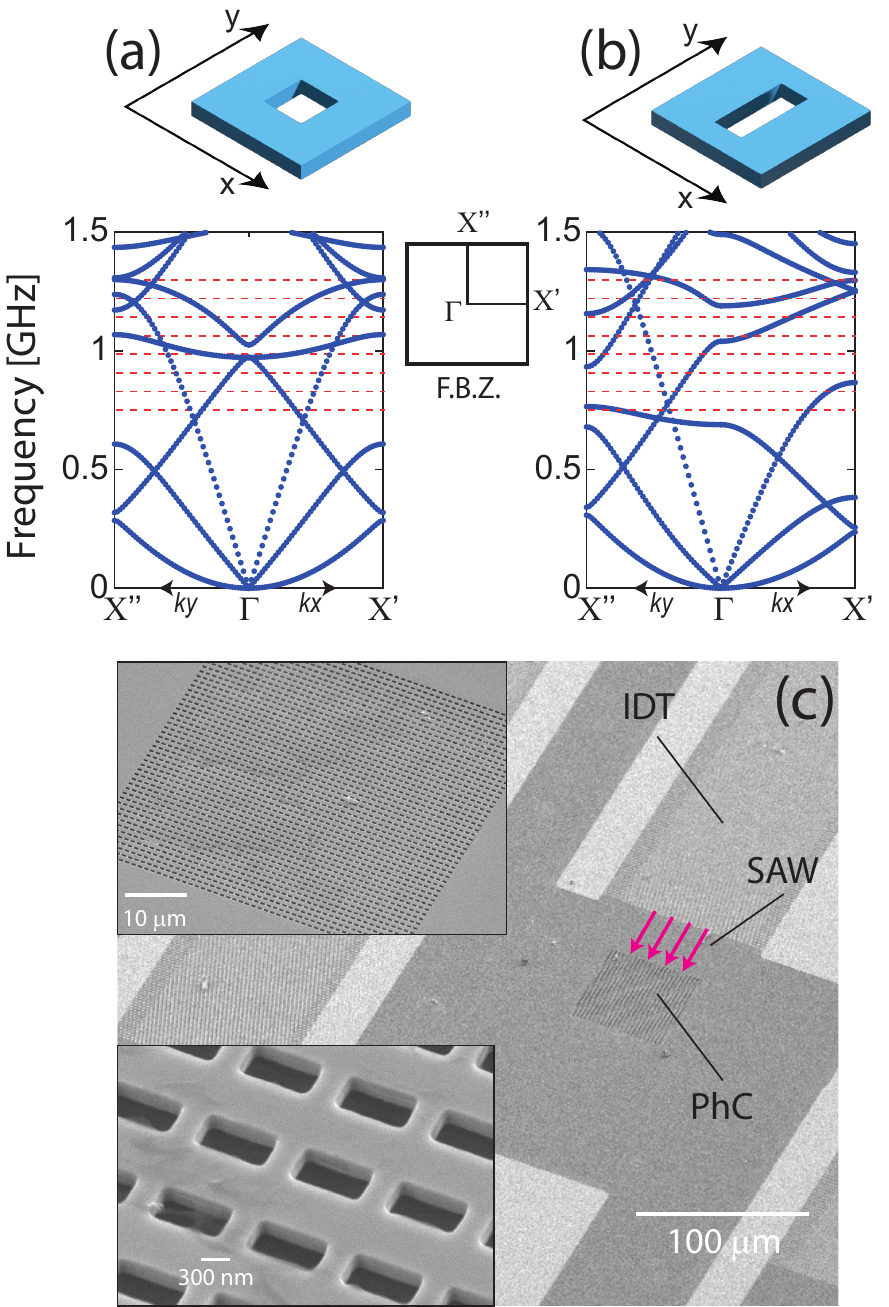}
\caption{(a-b): Sketch of the unit cell and simulated phononic band structure for the in-plane infinite PhCs consisting of square ($S_\mathrm{PhC}$) (a) and rectangular ($R_\mathrm{PhC}$) (b) holes etched in a 220~nm-thick GaAs membrane. The horizontal dashed lines indicate the resonant frequencies of the interdigital transducers (IDTs) for surface acoustic waves (SAWs) used to excite the $PhCs$. (c): Scanning electron micrograph (SEM) of one of the investigated device. The full released membrane and a zoom of few lattice periods are shown in the insets.}
\label{fig:1}
\end{figure}

We designed a set of devices defined by a square lattice of holes within a suspended 220-nm thick GaAs slab. The lattice constant of the periodic pattern was chosen to be $a$ = 1.125 $\mu$m, while two different kind of holes were considered: square holes with sides $l = 0.47\cdot a$ [$S_\mathrm{PhC}$, see Fig.~\ref{fig:1}(a)] or rectangular holes with sides $l_x = 0.7\cdot a$ and $l_y = 0.32\cdot a$ [$R_\mathrm{PhC}$, see Fig.~\ref{fig:1}(b)], respectively. The different hole geometry translates into a remarkable different aspect in their simulated phononic band structures. The simulations have been performed using a commercial Finite-Element Method solver, more details can be found in ref. \cite{Zanotto2022}. As expected, the band structure of the $R_\mathrm{PhC}$ shows an asymmetric dispersion along the high-symmetry paths in the First Brillouin Zone (FBZ) $\Gamma\rightarrow X'\equiv(\pi/a,0)$ and $\Gamma\rightarrow X''\equiv(0,\pi/a)$ (see inset in Fig. \ref{fig:1}), whereas the $S_\mathrm{PhC}$ bands are symmetric along the same paths.

\begin{figure*}[t!]
\centering
\includegraphics[width=0.85\textwidth]{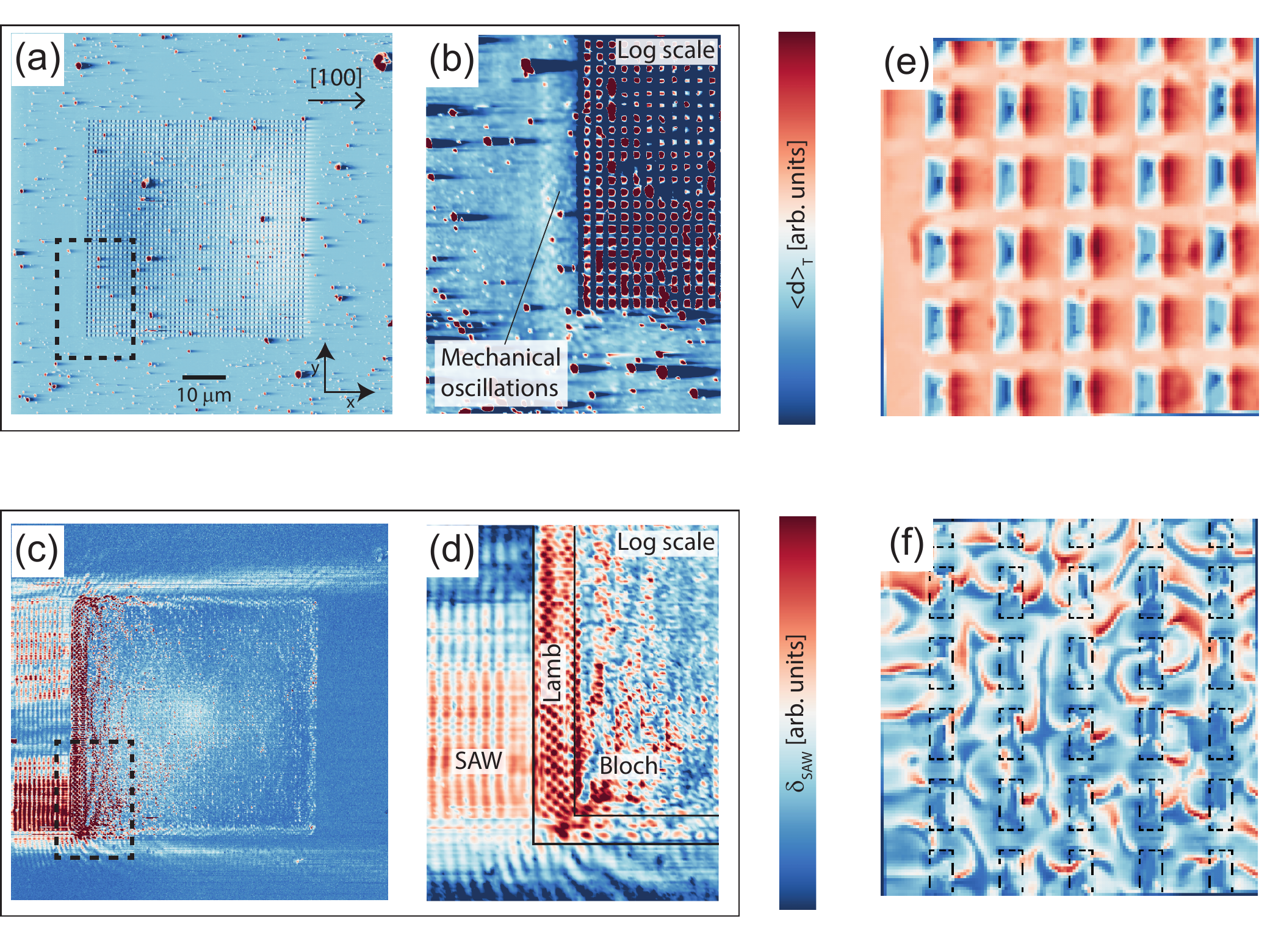}
\caption{Direct (i.e., $\deltaT$, cf. Eq.~\ref{eq:d_ave}) (a) and demodulated ($\deltaSAW$) (b) tip deflection signal detected on the $R_\mathrm{PhC}$ excited with a 1.262~GHz SAW, both in a linear color scale. (b) - (d): Magnified sections of each map in a logarithmic color scale. Few unit cells maps of direct (e) and demodulated (f) signals. In the latter map the $PhC$ holes have been superimposed as dashed rectangles.}
\label{fig:2}
\end{figure*}

The lowest mechanical band in Figs.~\ref{fig:1}(a) and \ref{fig:1}(a), characterized by a typical parabolic dispersion, is ultimately responsible for the effective parameters of the low frequency \textit{drum} modes, which are determined by the patterned region shape and size and are characterized by strong out-of-plane displacements. An effect of the asymmetric patterning is already present at these low frequency modes, which are  well within the metasurface regime (i.e., with all structural length scales smaller than the acoustic wavelength~\cite{Conte2022}). Conversely, here we investigate the role of  asymmetry  in the phononic crystal regime, where the structural length scales are comparable with the acoustic wavelength. To perform a spectroscopic analysis of the $PhC$ devices, we included eight kinds of interdigitated transducers (IDTs). Each kind of IDT was placed in front of one $PhC$ membrane of the two different ones shown in Fig. \ref{fig:1} (a) and (b). The different IDTs have been designed to resonantly excite Rayleigh surface acoustic waves at the different frequencies indicated by the dashed lines in Fig.~\ref{fig:1}(a) and (b). All single IDTs were composed by 150 metal finger pairs with an acoustic aperture of 180~$\mu$m, in such a way that they generate a SAW wavefront wider than the 150~$\mu$m lateral size of the PhC devices. More details on sample fabrication are reported in the Supplementary Information. A Scanning Electron Microscopy of a typical device is displayed in Fig.~\ref{fig:1} (c), showing the IDT and two magnifications of the $R_\mathrm{PhC}$ membrane. By driving the IDT at its resonant frequency, a SAW is generated  along [110] through the piezoelectricity of the GaAs layer; the generated SAW impinges on the PhC and therein excites mechanical modes that depend on the symmetry of the SAW field.

The mapping of the acoustic fields has been done employing the AAFM technique \cite{Rabe1994,Hesjedal2010}. The AFM detection tip, with resonance frequencies in the kHz range, does not directly respond to the GaAs high-frequency SAW vibrations; the technique then relies on the nonlinear interaction between the tip and the sample surface, which arises from a combination of Coulomb and Van der Waals interactions \cite{Sharahi2021}. Such a nonlinearity is well-manifested in the dependence of the force acting on the cantilever ($F$) on the tip-to-surface distance $z$ (see Supplementary Information). A simplified description of the system approximates the tip as a harmonic oscillator subjected to a time-dependent external forcing $F(z)$ induced by the distance modulation from a Rayleigh SAW, $z = z_0 + z_{SAW}$. Here, $z_{SAW}$ is a harmonic function representing an out-of-plane, standing wave displacement, i.e. $z_{SAW} = A cos(\omega_{SAW} t)$. In standard AFMs, the tip fastest response is in the 100s kHz range, limited by the mechanical resonances of the tip-hosting cantilever and the frequency response of the 4-quadrant photodetector used to probe the tip deflection. While this is significantly slower than GHz frequency of the SAWs, a displacement contribution appears in the time-averaged, `slow'   deflection $\langle d \rangle_T$, \cite{Hellemann2022}:
\begin{equation}
\label{eq:d_ave}
\deltaT =  \frac{1}{k} F(z) + \underbrace{\frac{1}{k} \left(F''(z) \frac{F'(z)}{F'(z)+k} + \frac{F''(z)}{2} \right)A^2}_{\deltaSAW},
\end{equation}
where $k$ is the tip cantilever spring constant and all the differentiations are with respect to the $z$ coordinate. Analyzing the right-hand side of Eq.~(\ref{eq:d_ave}), one notices a first linear term $F(z)/k$, which is responsible for the standard AFM signal yielding the sample topography. Additionally a nonlinear contribution proportional to a combination of the first and second derivative of $F$ and the SAW intensity $A^2$ appears. The derivation of Eq.~\ref{eq:d_ave} assumed a purely vertical tip-surface distance modulation, which we consider to be the most prominent component detected by AAFM. 
Note, however, that this equation can be obtained considering both travelling and standing acoustic fields. In the latter case, the spatial variations of the average acoustic amplitude enables the mapping of the acoustic wavefronts. In the former, in constrast, travelling phase information is lost and one obtains a signal proportional to the root-mean-square (RMS) of the squared amplitude of the acoustic field  $A^2$ , see Supplementary Information. 

The generally small signal proportional to $A^2$ can be efficiently detected by employing a lock-in amplifier (LIA). Here, the rf-signal driving the SAW is amplitude modulated at a frequency $f_m$ and the AFM signal component at this frequency detected by the LIA. The LIA detectioncan be maximized by choosing a modulation frequency coincident with one of the narrow, higher order tip cantilever modes. In our experiment, we used a modulation frequency of 536 kHz mode, which has shown to give the optimal overall results by minimizing the cross-talk between the acoustic displacement and the low frequency AFM topographic signal. More details on the experimental measurements can be found in the Supplementary Information.

\begin{figure}[t!]
\centering
\includegraphics[width=1\linewidth]{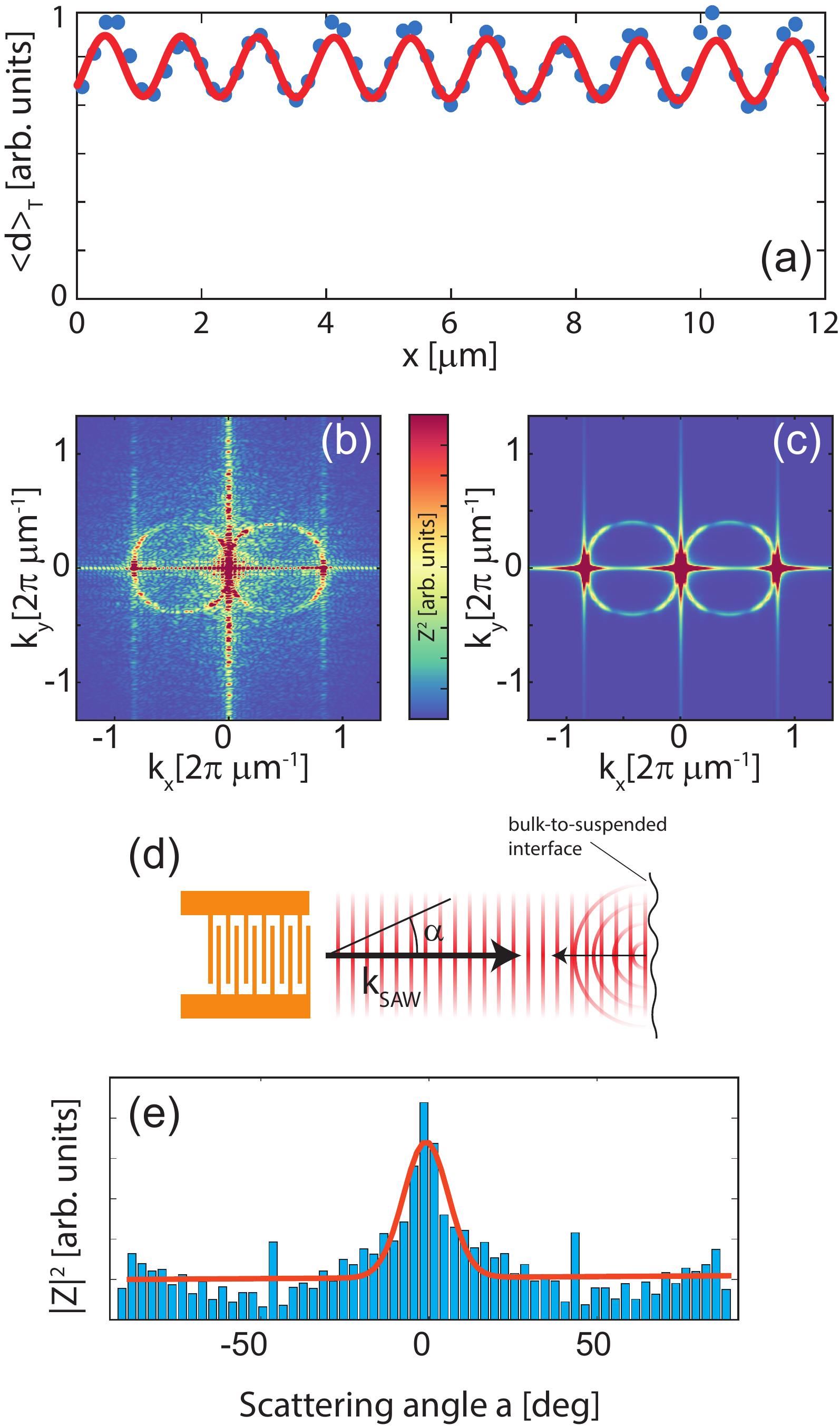}
\caption{(a): AAFM signal in the region between the IDT and the PhC integrated in the direction perpendicular to the SAW ($y$ direction). (b): bidimensional FT of the incoming SAW wave. (c): bidimensional FT of a toy-model considering a plane wave and an inhomogeneous cylindrical wave. (d): Sketch of the scattering configuration. (e): Angular spread of the wave impinging on the $R_\mathrm{PhC}$ device of Fig. \ref{fig:2}. The red line is the best Gaussian fit to the histogram.}
\label{fig:3}
\end{figure}

\section{Results}

\subsection{AAFM field mapping}

Figures~\ref{fig:2} (a) and (c) compare the directly acquired (i.e., $\deltaT$ in Eq.~\ref{eq:d_ave}) and the lock-in demodulated AFM deflection signal ($\deltaSAW$) for the $R_\mathrm{PhC}$ excited at 1.262 GHz by applying a 20~dBm rf-bias to the IDT (the IDT is placed on the left, outside the picture). As expected from Eq. (\ref{eq:d_ave}), the signal in panel (a) is essentially proportional to the surface topography: a clear square lattice made of rectangular holes is visible in the center of the map, as well as dust microparticles on the sample surface. Plotting the deflection map in a logarithmic scale allows one to better appreciate the feeble signal originating from mechanical vibrations, as indicated in the magnified view of Fig.~\ref{fig:2}(b).

By amplifying the $\delta_{SAW}$ term in Eq. (\ref{eq:d_ave}) through lock-in demodulation, we obtain the map reported in Fig. \ref{fig:2} (c). Here, the incoming SAW wave is clearly visible as well as several complex features in the patterned region. The magnification of the membrane lower left corner in panel (d) allows one to easily distinguish three different propagation regions: (i) from the Rayleigh SAW in the bulk, to (ii) a Lamb wave on the unpatterned part of the  slab to finally (iii) a Bloch wave inside the phononic crystal patterned in the slab. The mechanical waves can be clearly distinguished, although the high spatial resolution of the AAFM enables us to analyze even smaller features in the devices. 

As an example, Fig. \ref{fig:2} (e) shows the deflection map of a few phononic crystal unit cells with a resolution of roughly 50 nm. Considering the scanning parameters (i.e. large cantilever from surface distance) and high speed employed in our measurement, the tip can only partly feel the presence of a hole underneath it, resulting in a reduced, but not vanishing deflection signal in the hole regions. The corresponding displacement square map for the same area is shown in Fig.~\ref{fig:2} (f). As a visual aid, we have superimposed outlines of  the hole shapes on the map. Here, one can observe the complex displacement pattern resulting from the interference of multiple Bloch modes. 

\subsection{SAW scattering}

The experimental results of Fig.~\ref{fig:2} can be further analyzed to obtain information related to  acoustic wave scattering. At first, we focus on the portion of the sample between the IDT and the $PhC$ device. Here, the SAW propagates as an quasi-plane wave with wavevector along the $\hat{x}$ direction and a strong displacement component along $\hat{z}$. To evaluate the average SAW amplitude, we integrated the signal from Fig.~\ref{fig:2} (c) along $\hat{y}$; the result is displayed in Fig.~\ref{fig:3}(a). The experimental data can be fitted with a cosine square function, $f(x) = A + B \cos(k_0 x)^2$ (red curve). The obtained wavector is $k_0 = 2.56 \mu m^{-1}$, resulting in a SAW phase velocity of $\sim 3094\ m/s$, which is compatible with the value of $3022\ m/s$ determined from an elastic model for the sample structure,  see for example \cite{deLima2005_rev}. 

Additional information about scattering can be obtained by directly taking the 2D Fourier Transform (FT) of the entire area between the IDT and the phononic crystal. The reciprocal space map obtained by operating on Fig. \ref{fig:2} (c) is reported in Fig. \ref{fig:3} (b). It shows several interesting features: at first, we observe the three bright spots at $k_x=0$ and $k_x=\pm0.827\cdot2\pi\; \mu m^{-1} \approx \pm2 k_0$, respectively. These are expected by considering that the FT of a cosine squared function produces three $\delta$-functions according to:
\begin{equation}
FT[A + B cos(k_0 x)^2] \propto \delta(k_x) + \delta(k_x - 2 k_0) + \delta(k_x + 2 k_0).
\end{equation}      

In addition to the three bright spots, Fig.~\ref{fig:2}(b) also shows two symmetric, slightly deformed ring-shaped structures of radius $k_0$  centered at $[\pm k_0,0]$. These rings originate from the interference between the incoming SAW with waves non-specularly reflected at the  interface between bulk and suspended membrane or scattered at single  point-defects, see the scheme in Fig. \ref{fig:3} (d). As will be demonstrated later, the large mismatch in acoustic  impedance at the bulk-suspended membrane interface leads to strong reflections at this interface. In fact, all the features can be qualitatively reproduced by considering a numerical model combining a plane wave (generated SAW) with a weaker cylindrical wave plus an angular-dependent scattering amplitude (the scattered SAWs). Additionally, slightly anisotropic wave propagation has been considered by evaluating the slowness curve in our material heterostructure. The model is described in detail in the Supplementary Information. The simulated map is reported in Fig. \ref{fig:3} (c) and well agrees with the experimental one. Interestingly, the slight deformation of the ring structure that deviates from a perfect circle is well reproduced in the theoretical map and originates from the material anisotropy.

The wave dispersion due to scattering can be more quantitatively estimated from the map of Fig.~\ref{fig:3}(b). Panel (e) of Fig.~\ref{fig:3} is derived from the reciprocal map as a histogram of the FT-power as a function of the polar angle, considering $\hat{k_x}$ as the zero-angle direction, as sketched in the figure. As can be seen, the histogram has a maximum along the planar SAW propagation direction along $\hat{k_x}$. The maximum then slowly decreases towards a weaker and constant backgroud, which spans from the weak scattering at large angles due to random particles and defects. The histogram of Fig. \ref{fig:3} (c) can be fitted with an offsetted Gaussian function, resulting in an angular spread $\Delta \alpha = 9.6^\circ\pm2.5^\circ$, for the device under investigation.

\begin{figure}[h!]
\centering
\includegraphics[width=1\linewidth]{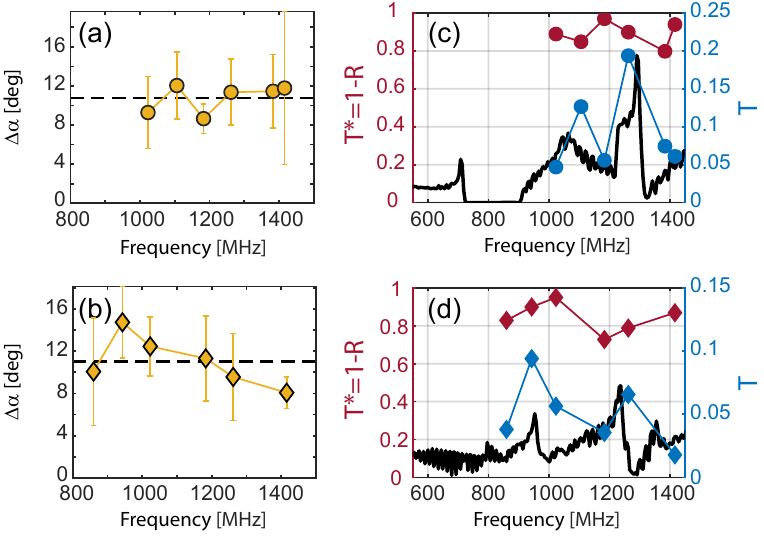}
\caption{Left column: angular dispersion of the incoming SAW evaluated from AAFM reciprocal space maps for $R_\mathrm{PhC}$ (a) and $S_\mathrm{PhC}$ (b). Deduced transmissitivy from the standing to travelling wave ratio (red) and directly evaluated transmissivity (blue) are compared with the crystal DOS (black) for $R_\mathrm{PhC}$ (c) and $S_\mathrm{PhC}$ (d).}
\label{fig:4}
\end{figure}

A similar analysis was applied to the whole set of devices comprising 6 $S_\mathrm{PhC}$ and 6 $R_\mathrm{PhC}$ phononic crystals excited at the 8 frequencies, ranging from $\sim$850 MHz to $\sim$1.42 GHz, indicated by the dashed lines in Fig.~\ref{fig:1}(a) and \ref{fig:1}(b). The results are summarized in  panels (a) and (b) of Fig.~\ref{fig:4}, respectively. The angular spread $\Delta \alpha$ is, to a good approximation,  rougly constant for all the investigated devices with  average values of  $\sim 10.8^\circ$ and $\sim 11^\circ$ for $S_\mathrm{PhC}$ and $R_\mathrm{PhC}$, respectively. This observation strengthens the hypotesis that the SAW angular spread mainly originates at the bulk-to-suspended interface and it is, therefore, independent from the structure of the $PhC$ fabricated within the membrane. In fact,  scattering events induced by the $PhC$ structure  would carry the signatures of its non-trivial energy band dispersion illustrated in Figs.~\ref{fig:1}(a) and \ref{fig:1}(b).

Looking for a signature in the wave spectral scattering, we used our AAFM characterization to evaluate the reflectivity (R) and trasmissivity (T) of acoustic waves through the $PhC$ membrane. 
An estimation of the reflectivity can be performed by observing that the incoming SAW integrated square displacement in Fig. \ref{fig:3} (a) is composed by an oscillating term on constant background. Our nonlinear detection mechanism averages the signal to a timescale which is orders of magnitude slower than the SAW period. While this is not an issue for standing waves, which have fixed nodes in the spatial coordinates, it produces a constant signal for travelling waves, whose crests travels faster than the detection integration time. Specifically, this would correspond to the squared wave RMS amplitude value. These observations suggest that the incoming SAW will likely be composed by a standing and a travelling wave components, as expected for a partially reflected wave. With a simple analytical, one-dimensional model, it is possible to evaluate the SAW reflected amplitude, see Suppletementary Information for more details. This analysis results in a discrete spectral reflectivity $R(f_{IDT})$, evaluated at the frequency of each IDT for the two samples geometry. To correlate the estimated reflectivity with the $PhC$-induced scattering we can consider its deduced transmissivity, $T^*(f_{IDT})=1-R(f_{IDT})$, which is a valid approximation if one neglects internal device losses. The deduced transmissivity is plotted in Fig. \ref{fig:4} (c) and (d), for $R_\mathrm{PhC}$ and $S_\mathrm{PhC}$, respectively. In a rough approximation neglecting the coupling between SAW and $PhC$ Bloch modes, the deduced transmissivity can be compared with the integrated $PhC$ density of states (DOS). As can be seen, $T^*$ poorly correlates with the DOS, once more confirming that the wave reflectivity does not depend on the local $PhC$ device, but rather on the bulk-to-membrane interface.

Once the reflectivity has been assessed, the transmissivity can be easily obtained by dividing the integrated total AAFM signal in two planar regions placed before and after the membrane, opportunely modified to consider the (spurious) reflections previously described. The final results for $T$ are added as blue symbols to Fig.~\ref{fig:4} (c) and (d) and show a good correlation with the DOS, for both devices geometries. This suggests that the $PhC$ can have a strong impact on the propagating SAW and, therefore, can be used for wave manipulation in the GHz frequency range employing subwavelength thick suspended membranes.\\
\begin{figure*}[t!]
\centering
\includegraphics[width=0.85\textwidth]{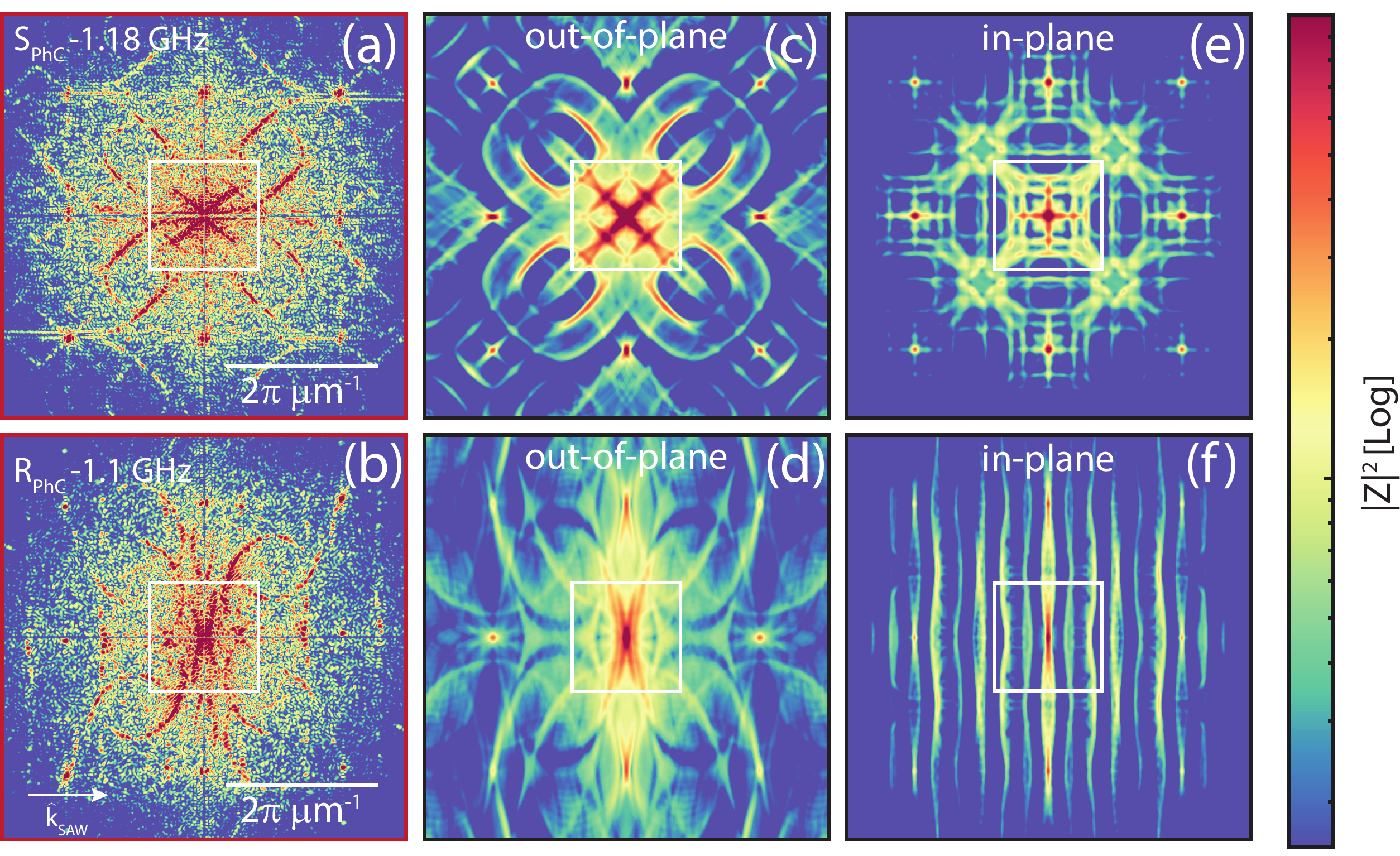}
\caption{(a)-(b): experimental reciprocal space maps of the AAFM signal within the patterned region. Selected maps coming of the $S_\mathrm{PhC}$ (a) and $R_\mathrm{PhC}$ (b) have been considered. (c)-(f): simulated maps for the same $PhC$ in a color scale considering the out-of-plane (c-d) and in-plane (e-f) displacement components, respectively. The white square represents the edge of the First Brillouin Zone.}
\label{fig:5}
\end{figure*}

\subsection{Reciprocal space mapping of phononic crystals}

Additional insights into the $PhC$ effect on wave manipulation can be obtained by inspecting the AAFM signal within the membrane region. While the experimental maps in the real space can be of difficult interpretation, their two-dimensional FT can highlight interesting features  linked to beam steering and ordinary or negative wave refraction \cite{He2018,Zanotto2022}. In our device configuration, the SAW impinges on the $PhC$ with a set of wavevectors ($|k_{SAW}|=2\pi f_{IDT}/v_s$, $v_s$ being the sound velocity) broadened by the angular spread reported in panel (a) and (b) of Fig. \ref{fig:3}. Therefore, one would expect to excite only a certain region of the reciprocal space, namely around $\pm k_{SAW}$. Surprisingly, as can be inspected by the selected experimental maps of Fig. \ref{fig:5} (a) and (b), a very wide region of reciprocal space is populated, even in the direction perpendicular to $\hat{k}_{SAW}$. This suggests strong wave scattering within the $PhC$ itself, which can originate both from random disorder and/or governed by the hole symmetry.\\ 
We note that several detailed features are present in the experimental maps, allowing for an in-depth investigation of the $PhC$ eigenmodes as well as of the $PhC$ wave manipulation properties. The net effect of the pattern can be evaluated by comparing the experimental reciprocal space maps with FEM simulations, see Fig. \ref{fig:5}. The simulated maps have been evaluated starting from the isofrequency contours in the reciprocal space, obtained from a slice of the 2D band structure simulations. Its self-convolution integral is then equivalent to the FT of the squared signal, which is the quantity measured by AAFM technique. Moreover, the simulated maps have been colored using a logarithmic weight proportional to the out-of-plane, $\hat{z}$ and in plane,$\hat{x}-\hat{y}$ displacement of the vibrations, more details can be found in \cite{Zanotto2022}. As a general observation, we found an excellent agreement between experiments and out-of-plane simulations, with fine features clearly present in both maps. Conversely, we do not see any relevant correlation between experiments and in-plane simulations, suggesting that in our measurement configuration the AAFM is mostly sensitive to the out-of-plane mechanical vibrations. Indeed, the AAFM can be tailored for an enhanced sensitivity in the detection of planar modes, especially when in proximity to a hole \cite{Behme2001}, although this feature is not relevant in this experiment. The distortions and different amplitude of some of the experimental features with respect to the simulated ones can be imputed to the non-flatness of the membranes, which can be slightly buckled due to the GaAs material softness. 
Another interesting observation can be done by considering the general shape of the reciprocal space maps, which is more isotropic for the $S_\mathrm{PhC}$ (panels a and c), while more elongated for the $R_\mathrm{PhC}$ (panels b and d), respectively. This is linked to the pattern symmetry, which enhances or suppresses wave scattering in certain direction according to the energy-momentum dispersion of the Bloch modes. A more quantitive analysis about the reciprocal maps isotropy has been performed by comparing the distribution of modes along $\hat{k}_x$ and $\hat{k}_y$ directions. $S_\mathrm{PhC}$ shows very similar distributions, with an $\Delta k_x$ to $\Delta k_y$ average ratio of $0.96\pm0.08$; conversely, $R_\mathrm{PhC}$ shows an average ratio of $0.72\pm0.08$, with a broader distribution $\Delta k_y$ given by the reduced symmetry of its unit cell. Details on the reciprocal space isotropy analysis are available in the Supplementary Information.\\ 
The results here illustrated show that, on top of an inhomogeneous scattering background due to pervasively present device non-idealities, a strong phonon scattering channel is governed by the local band structure, which ultimately determines the density of available states in the system. In analogy with transport in optical system, one can imagine several ways for manipulating elastic waves by density of states engineering. As an example, one can use $S_\mathrm{PhC}$ as complex spectral filters, while $R_\mathrm{PhC}$  grants an additional spatial dependent effect, stopping waves propagating along a certain direction while allowing the transmission of wave coming, for example, along the perpendicular one. This and more complex features can be introduced to create complex nodes in elastic networks, which can act as manipulators for microwave-based telecom technology.

\section{Conclusions}

In this manuscript, we have introduced an AFM-based technique for the  detailed characterization of GHz phononic crystal membrane devices. The high spatial resolution of this technique along with its rapid scanning times makes it perfectly suitable for investigating mechanical devices operating at GHz frequencies and above. Elastic photonic crystal and mechanical metasurfaces supporting high frequency waves can find a wide application in the next-generation telecommunication technologies, where electromagnetic waves can be processed by coupling them to mechanical channels, as routinely done with SAW-filter in mobile device technology. The design flexibility of periodic devices opens the way for complex wave-based functionalities, possibly enabling the emergence of 6G technologies, where the increased carrier wave frequency could be accompanied by parallel manipulation and multiplexing based on spatial, spectral and/or polarization features. As a proof-of-concept, here we have studied  with the AAFM technique a set of phononic crystal with different symmetries and operating at $\sim$ 1 GHz frequency. By employing reciprocal space analysis we have quantified the role of inhomogeneous and homogeneous scattering, the latter showing features connected with the crystal symmetry itself. The simple geometrical features employed here can be useful for realizing complex spectral filters which are additionally dependent on the elastic wavevectors.

\medskip
\textbf{Supporting Information} \par 
Supporting Information is available from the Wiley Online Library or from the author.

\medskip
\textbf{Acknowledgements} \par 
We acknowledge Dr. Nazim Ashurbekov for a critical reading of the manuscript. Alexander von Humboldt Foundation is gratefully acknowledged for funding this research through the Experienced Researcher Fellowship program.

\medskip

\bibliography{AFM_SAW.bib}

\begin{thebibliography}{36}%
\makeatletter
\providecommand \@ifxundefined [1]{%
 \@ifx{#1\undefined}
}%
\providecommand \@ifnum [1]{%
 \ifnum #1\expandafter \@firstoftwo
 \else \expandafter \@secondoftwo
 \fi
}%
\providecommand \@ifx [1]{%
 \ifx #1\expandafter \@firstoftwo
 \else \expandafter \@secondoftwo
 \fi
}%
\providecommand \natexlab [1]{#1}%
\providecommand \enquote  [1]{``#1''}%
\providecommand \bibnamefont  [1]{#1}%
\providecommand \bibfnamefont [1]{#1}%
\providecommand \citenamefont [1]{#1}%
\providecommand \href@noop [0]{\@secondoftwo}%
\providecommand \href [0]{\begingroup \@sanitize@url \@href}%
\providecommand \@href[1]{\@@startlink{#1}\@@href}%
\providecommand \@@href[1]{\endgroup#1\@@endlink}%
\providecommand \@sanitize@url [0]{\catcode `\\12\catcode `\$12\catcode
  `\&12\catcode `\#12\catcode `\^12\catcode `\_12\catcode `\%12\relax}%
\providecommand \@@startlink[1]{}%
\providecommand \@@endlink[0]{}%
\providecommand \url  [0]{\begingroup\@sanitize@url \@url }%
\providecommand \@url [1]{\endgroup\@href {#1}{\urlprefix }}%
\providecommand \urlprefix  [0]{URL }%
\providecommand \Eprint [0]{\href }%
\providecommand \doibase [0]{https://doi.org/}%
\providecommand \selectlanguage [0]{\@gobble}%
\providecommand \bibinfo  [0]{\@secondoftwo}%
\providecommand \bibfield  [0]{\@secondoftwo}%
\providecommand \translation [1]{[#1]}%
\providecommand \BibitemOpen [0]{}%
\providecommand \bibitemStop [0]{}%
\providecommand \bibitemNoStop [0]{.\EOS\space}%
\providecommand \EOS [0]{\spacefactor3000\relax}%
\providecommand \BibitemShut  [1]{\csname bibitem#1\endcsname}%
\let\auto@bib@innerbib\@empty
\bibitem [{\citenamefont {Rayleigh}(1885)}]{Rayleigh1885}%
  \BibitemOpen
  \bibfield  {author} {\bibinfo {author} {\bibfnamefont {L.}~\bibnamefont
  {Rayleigh}},\ }\bibfield  {title} {\bibinfo {title} {{On Waves Propagated
  along the Plane Surface of an Elastic Solid}},\ }\href
  {https://doi.org/10.1112/plms/s1-17.1.4} {\bibfield  {journal} {\bibinfo
  {journal} {Proceedings of the London Mathematical Society}\ }\textbf
  {\bibinfo {volume} {s1-17}},\ \bibinfo {pages} {4} (\bibinfo {year}
  {1885})},\ \Eprint
  {https://arxiv.org/abs/https://academic.oup.com/plms/article-pdf/s1-17/1/4/4368144/s1-17-1-4.pdf}
  {https://academic.oup.com/plms/article-pdf/s1-17/1/4/4368144/s1-17-1-4.pdf}
  \BibitemShut {NoStop}%
\bibitem [{\citenamefont {Reza~Ali}\ and\ \citenamefont
  {Prasad}(2020)}]{RezaAli2020}%
  \BibitemOpen
  \bibfield  {author} {\bibinfo {author} {\bibfnamefont {W.}~\bibnamefont
  {Reza~Ali}}\ and\ \bibinfo {author} {\bibfnamefont {M.}~\bibnamefont
  {Prasad}},\ }\bibfield  {title} {\bibinfo {title} {Piezoelectric mems based
  acoustic sensors: A review},\ }\href@noop {} {\bibfield  {journal} {\bibinfo
  {journal} {Phys. Rev. Appl.}\ }\textbf {\bibinfo {volume} {301}},\ \bibinfo
  {pages} {111756} (\bibinfo {year} {2020})}\BibitemShut {NoStop}%
\bibitem [{\citenamefont {Devkota}\ \emph {et~al.}(2017)\citenamefont
  {Devkota}, \citenamefont {Ohodnicki},\ and\ \citenamefont
  {Greve}}]{Devkota2017}%
  \BibitemOpen
  \bibfield  {author} {\bibinfo {author} {\bibfnamefont {J.}~\bibnamefont
  {Devkota}}, \bibinfo {author} {\bibfnamefont {P.~R.}\ \bibnamefont
  {Ohodnicki}},\ and\ \bibinfo {author} {\bibfnamefont {D.~W.}\ \bibnamefont
  {Greve}},\ }\bibfield  {title} {\bibinfo {title} {Saw sensors for chemical
  vapors and gases},\ }\href@noop {} {\bibfield  {journal} {\bibinfo  {journal}
  {Sensors}\ }\textbf {\bibinfo {volume} {17}},\ \bibinfo {pages} {801}
  (\bibinfo {year} {2017})}\BibitemShut {NoStop}%
\bibitem [{\citenamefont {L\"{a}nge}\ \emph {et~al.}(2008)\citenamefont
  {L\"{a}nge}, \citenamefont {Rapp},\ and\ \citenamefont {Rapp}}]{Lange2008}%
  \BibitemOpen
  \bibfield  {author} {\bibinfo {author} {\bibfnamefont {K.}~\bibnamefont
  {L\"{a}nge}}, \bibinfo {author} {\bibfnamefont {B.~E.}\ \bibnamefont
  {Rapp}},\ and\ \bibinfo {author} {\bibfnamefont {M.}~\bibnamefont {Rapp}},\
  }\bibfield  {title} {\bibinfo {title} {Surface acoustic wave biosensors: a
  review},\ }\href@noop {} {\bibfield  {journal} {\bibinfo  {journal} {Anal.
  Bioanal. Chem.}\ }\textbf {\bibinfo {volume} {391}},\ \bibinfo {pages} {1509}
  (\bibinfo {year} {2008})}\BibitemShut {NoStop}%
\bibitem [{\citenamefont {Morgan}(2010)}]{Morgan2010}%
  \BibitemOpen
  \bibfield  {author} {\bibinfo {author} {\bibfnamefont {D.}~\bibnamefont
  {Morgan}},\ }\href@noop {} {\emph {\bibinfo {title} {Surface acoustic wave
  filters: With applications to electronic communications and signal
  processing}}}\ (\bibinfo  {publisher} {Academic Press},\ \bibinfo {year}
  {2010})\BibitemShut {NoStop}%
\bibitem [{\citenamefont {Bernardo}(2002)}]{Bernardo2002}%
  \BibitemOpen
  \bibfield  {author} {\bibinfo {author} {\bibfnamefont {R.~P.}\ \bibnamefont
  {Bernardo}},\ }\bibfield  {title} {\bibinfo {title} {Saw voltage-controlled
  oscillators.(application note)},\ }\href@noop {} {\bibfield  {journal}
  {\bibinfo  {journal} {Microwave Journal}\ }\textbf {\bibinfo {volume} {45}},\
  \bibinfo {pages} {166} (\bibinfo {year} {2002})}\BibitemShut {NoStop}%
\bibitem [{\citenamefont {Hashimoto}(2000)}]{Hashimoto2000}%
  \BibitemOpen
  \bibfield  {author} {\bibinfo {author} {\bibfnamefont {K.-Y.}\ \bibnamefont
  {Hashimoto}},\ }\href@noop {} {\emph {\bibinfo {title} {Surface Acoustic Wave
  Devices in Telecommunications}}}\ (\bibinfo  {publisher} {Springer},\
  \bibinfo {year} {2000})\BibitemShut {NoStop}%
\bibitem [{\citenamefont {Delsing}\ \emph {et~al.}(2019)\citenamefont
  {Delsing}, \citenamefont {Cleland}, \citenamefont {Schuetz}, \citenamefont
  {Kn{\"o}rzer}, \citenamefont {Giedke}, \citenamefont {Cirac}, \citenamefont
  {Srinivasan}, \citenamefont {Wu}, \citenamefont {Balram}, \citenamefont
  {B{\"a}uerle} \emph {et~al.}}]{Delsing2019}%
  \BibitemOpen
  \bibfield  {author} {\bibinfo {author} {\bibfnamefont {P.}~\bibnamefont
  {Delsing}}, \bibinfo {author} {\bibfnamefont {A.~N.}\ \bibnamefont
  {Cleland}}, \bibinfo {author} {\bibfnamefont {M.~J.}\ \bibnamefont
  {Schuetz}}, \bibinfo {author} {\bibfnamefont {J.}~\bibnamefont
  {Kn{\"o}rzer}}, \bibinfo {author} {\bibfnamefont {G.}~\bibnamefont {Giedke}},
  \bibinfo {author} {\bibfnamefont {J.~I.}\ \bibnamefont {Cirac}}, \bibinfo
  {author} {\bibfnamefont {K.}~\bibnamefont {Srinivasan}}, \bibinfo {author}
  {\bibfnamefont {M.}~\bibnamefont {Wu}}, \bibinfo {author} {\bibfnamefont
  {K.~C.}\ \bibnamefont {Balram}}, \bibinfo {author} {\bibfnamefont
  {C.}~\bibnamefont {B{\"a}uerle}}, \emph {et~al.},\ }\bibfield  {title}
  {\bibinfo {title} {The 2019 surface acoustic waves roadmap},\ }\href@noop {}
  {\bibfield  {journal} {\bibinfo  {journal} {Journal of Physics D: Applied
  Physics}\ }\textbf {\bibinfo {volume} {52}},\ \bibinfo {pages} {353001}
  (\bibinfo {year} {2019})}\BibitemShut {NoStop}%
\bibitem [{\citenamefont {Xu}\ \emph {et~al.}(2022)\citenamefont {Xu},
  \citenamefont {Harley}, \citenamefont {Ma}, \citenamefont {Lee},\ and\
  \citenamefont {Collins}}]{Xu2022}%
  \BibitemOpen
  \bibfield  {author} {\bibinfo {author} {\bibfnamefont {M.}~\bibnamefont
  {Xu}}, \bibinfo {author} {\bibfnamefont {W.~S.}\ \bibnamefont {Harley}},
  \bibinfo {author} {\bibfnamefont {Z.}~\bibnamefont {Ma}}, \bibinfo {author}
  {\bibfnamefont {P.~V.~S.}\ \bibnamefont {Lee}},\ and\ \bibinfo {author}
  {\bibfnamefont {D.~J.}\ \bibnamefont {Collins}},\ }\bibfield  {title}
  {\bibinfo {title} {Sound-speed modifying acoustic metasurfaces for acoustic
  holography},\ }\href@noop {} {\bibfield  {journal} {\bibinfo  {journal}
  {Advanced Materials}\ }\textbf {\bibinfo {volume} {n/a}},\ \bibinfo {pages}
  {2208002} (\bibinfo {year} {2022})}\BibitemShut {NoStop}%
\bibitem [{\citenamefont {He}\ \emph {et~al.}(2018)\citenamefont {He},
  \citenamefont {Qiu}, \citenamefont {Ye}, \citenamefont {Cai}, \citenamefont
  {Fan}, \citenamefont {Ke}, \citenamefont {Zhang},\ and\ \citenamefont
  {Liu}}]{He2018}%
  \BibitemOpen
  \bibfield  {author} {\bibinfo {author} {\bibfnamefont {H.}~\bibnamefont
  {He}}, \bibinfo {author} {\bibfnamefont {C.}~\bibnamefont {Qiu}}, \bibinfo
  {author} {\bibfnamefont {L.}~\bibnamefont {Ye}}, \bibinfo {author}
  {\bibfnamefont {X.}~\bibnamefont {Cai}}, \bibinfo {author} {\bibfnamefont
  {X.}~\bibnamefont {Fan}}, \bibinfo {author} {\bibfnamefont {M.}~\bibnamefont
  {Ke}}, \bibinfo {author} {\bibfnamefont {F.}~\bibnamefont {Zhang}},\ and\
  \bibinfo {author} {\bibfnamefont {Z.}~\bibnamefont {Liu}},\ }\bibfield
  {title} {\bibinfo {title} {Topological negative refraction of surface
  acoustic waves in a weyl phononic crystal},\ }\href@noop {} {\bibfield
  {journal} {\bibinfo  {journal} {Nature}\ }\textbf {\bibinfo {volume} {560}},\
  \bibinfo {pages} {61} (\bibinfo {year} {2018})}\BibitemShut {NoStop}%
\bibitem [{\citenamefont {Cha}\ \emph {et~al.}(2018)\citenamefont {Cha},
  \citenamefont {Kim},\ and\ \citenamefont {Daraio}}]{Cha2018}%
  \BibitemOpen
  \bibfield  {author} {\bibinfo {author} {\bibfnamefont {J.}~\bibnamefont
  {Cha}}, \bibinfo {author} {\bibfnamefont {K.}~\bibnamefont {Kim}},\ and\
  \bibinfo {author} {\bibfnamefont {C.}~\bibnamefont {Daraio}},\ }\bibfield
  {title} {\bibinfo {title} {Experimental realization of on-chip topological
  nanoelectromechanical metamaterials},\ }\href@noop {} {\bibfield  {journal}
  {\bibinfo  {journal} {Nature}\ }\textbf {\bibinfo {volume} {564}},\ \bibinfo
  {pages} {229} (\bibinfo {year} {2018})}\BibitemShut {NoStop}%
\bibitem [{\citenamefont {Zhang}\ \emph {et~al.}(2018)\citenamefont {Zhang},
  \citenamefont {Xiao}, \citenamefont {Cheng}, \citenamefont {Lu},\ and\
  \citenamefont {Christensen}}]{Zhang2018}%
  \BibitemOpen
  \bibfield  {author} {\bibinfo {author} {\bibfnamefont {X.}~\bibnamefont
  {Zhang}}, \bibinfo {author} {\bibfnamefont {M.}~\bibnamefont {Xiao}},
  \bibinfo {author} {\bibfnamefont {Y.}~\bibnamefont {Cheng}}, \bibinfo
  {author} {\bibfnamefont {M.-H.}\ \bibnamefont {Lu}},\ and\ \bibinfo {author}
  {\bibfnamefont {J.}~\bibnamefont {Christensen}},\ }\bibfield  {title}
  {\bibinfo {title} {Topological sound},\ }\href@noop {} {\bibfield  {journal}
  {\bibinfo  {journal} {Communication Physics}\ }\textbf {\bibinfo {volume}
  {1}},\ \bibinfo {pages} {97} (\bibinfo {year} {2018})}\BibitemShut {NoStop}%
\bibitem [{\citenamefont {Luo}\ \emph {et~al.}(2021)\citenamefont {Luo},
  \citenamefont {Wang}, \citenamefont {Lin}, \citenamefont {Jiang},
  \citenamefont {Wu}, \citenamefont {Li},\ and\ \citenamefont
  {J.-H.}}]{Luo2021}%
  \BibitemOpen
  \bibfield  {author} {\bibinfo {author} {\bibfnamefont {L.}~\bibnamefont
  {Luo}}, \bibinfo {author} {\bibfnamefont {H.-X.}\ \bibnamefont {Wang}},
  \bibinfo {author} {\bibfnamefont {Z.-K.}\ \bibnamefont {Lin}}, \bibinfo
  {author} {\bibfnamefont {B.}~\bibnamefont {Jiang}}, \bibinfo {author}
  {\bibfnamefont {Y.}~\bibnamefont {Wu}}, \bibinfo {author} {\bibfnamefont
  {F.}~\bibnamefont {Li}},\ and\ \bibinfo {author} {\bibfnamefont
  {J.}~\bibnamefont {J.-H.}},\ }\bibfield  {title} {\bibinfo {title}
  {Observation of a phononic higher-order weyl semimetal},\ }\href@noop {}
  {\bibfield  {journal} {\bibinfo  {journal} {Nature Materials}\ }\textbf
  {\bibinfo {volume} {20}},\ \bibinfo {pages} {794} (\bibinfo {year}
  {2021})}\BibitemShut {NoStop}%
\bibitem [{\citenamefont {Shao}\ \emph {et~al.}(2020)\citenamefont {Shao},
  \citenamefont {Mao}, \citenamefont {Maity}, \citenamefont {Sinclair},
  \citenamefont {Hu.}, \citenamefont {Lang},\ and\ \citenamefont
  {Lon\v{c}ar}}]{Shao2020}%
  \BibitemOpen
  \bibfield  {author} {\bibinfo {author} {\bibfnamefont {L.}~\bibnamefont
  {Shao}}, \bibinfo {author} {\bibfnamefont {W.}~\bibnamefont {Mao}}, \bibinfo
  {author} {\bibfnamefont {S.}~\bibnamefont {Maity}}, \bibinfo {author}
  {\bibfnamefont {N.}~\bibnamefont {Sinclair}}, \bibinfo {author}
  {\bibfnamefont {Y.}~\bibnamefont {Hu.}}, \bibinfo {author} {\bibfnamefont
  {Y.}~\bibnamefont {Lang}},\ and\ \bibinfo {author} {\bibfnamefont
  {M.}~\bibnamefont {Lon\v{c}ar}},\ }\bibfield  {title} {\bibinfo {title}
  {Non-reciprocal transmission of microwave acoustic waves in nonlinear
  parity–time symmetric resonators},\ }\href@noop {} {\bibfield  {journal}
  {\bibinfo  {journal} {Nat. Electron.}\ }\textbf {\bibinfo {volume} {3}},\
  \bibinfo {pages} {267} (\bibinfo {year} {2020})}\BibitemShut {NoStop}%
\bibitem [{\citenamefont {Gao}\ \emph {et~al.}(2023)\citenamefont {Gao},
  \citenamefont {Benchabane}, \citenamefont {Bermak}, \citenamefont {Dong},\
  and\ \citenamefont {Khelif}}]{Gao2023}%
  \BibitemOpen
  \bibfield  {author} {\bibinfo {author} {\bibfnamefont {F.}~\bibnamefont
  {Gao}}, \bibinfo {author} {\bibfnamefont {S.}~\bibnamefont {Benchabane}},
  \bibinfo {author} {\bibfnamefont {A.}~\bibnamefont {Bermak}}, \bibinfo
  {author} {\bibfnamefont {S.}~\bibnamefont {Dong}},\ and\ \bibinfo {author}
  {\bibfnamefont {A.}~\bibnamefont {Khelif}},\ }\bibfield  {title} {\bibinfo
  {title} {On-chip tightly confined guiding and splitting of surface acoustic
  waves using line defects in phononic crystals},\ }\href@noop {} {\bibfield
  {journal} {\bibinfo  {journal} {Adv. Funct. Mat.}\ }\textbf {\bibinfo
  {volume} {33}},\ \bibinfo {pages} {2213625} (\bibinfo {year}
  {2023})}\BibitemShut {NoStop}%
\bibitem [{\citenamefont {Zanotto}\ \emph {et~al.}(2019)\citenamefont
  {Zanotto}, \citenamefont {Biasiol}, \citenamefont {Santos},\ and\
  \citenamefont {Pitanti}}]{Zanotto2022}%
  \BibitemOpen
  \bibfield  {author} {\bibinfo {author} {\bibfnamefont {S.}~\bibnamefont
  {Zanotto}}, \bibinfo {author} {\bibfnamefont {G.}~\bibnamefont {Biasiol}},
  \bibinfo {author} {\bibfnamefont {P.}~\bibnamefont {Santos}},\ and\ \bibinfo
  {author} {\bibfnamefont {A.}~\bibnamefont {Pitanti}},\ }\bibfield  {title}
  {\bibinfo {title} {Metamaterial-enabled asymmetric negative refraction of ghz
  mechanical waves},\ }\href@noop {} {\bibfield  {journal} {\bibinfo  {journal}
  {Nature Communications}\ }\textbf {\bibinfo {volume} {13}},\ \bibinfo {pages}
  {5939} (\bibinfo {year} {2019})}\BibitemShut {NoStop}%
\bibitem [{\citenamefont {De~Lima~Jr.}\ and\ \citenamefont
  {Santos}(2005)}]{deLima2005}%
  \BibitemOpen
  \bibfield  {author} {\bibinfo {author} {\bibfnamefont {M.~M.}\ \bibnamefont
  {De~Lima~Jr.}}\ and\ \bibinfo {author} {\bibfnamefont {P.~V.}\ \bibnamefont
  {Santos}},\ }\bibfield  {title} {\bibinfo {title} {Modulation of photonic
  structures by surface acoustic waves},\ }\href@noop {} {\bibfield  {journal}
  {\bibinfo  {journal} {Rep. Prog. Phys.}\ }\textbf {\bibinfo {volume} {68}},\
  \bibinfo {pages} {7} (\bibinfo {year} {2005})}\BibitemShut {NoStop}%
\bibitem [{\citenamefont {Pitanti}\ \emph {et~al.}(2020)\citenamefont
  {Pitanti}, \citenamefont {Makkonen}, \citenamefont {Colombano}, \citenamefont
  {Zanotto}, \citenamefont {Vicarelli}, \citenamefont {Cecchini}, \citenamefont
  {Griol}, \citenamefont {Navarro-Urrios}, \citenamefont {Sotomayor-Torres},
  \citenamefont {Martinez},\ and\ \citenamefont {Ahopelto}}]{Pitanti2020}%
  \BibitemOpen
  \bibfield  {author} {\bibinfo {author} {\bibfnamefont {A.}~\bibnamefont
  {Pitanti}}, \bibinfo {author} {\bibfnamefont {T.}~\bibnamefont {Makkonen}},
  \bibinfo {author} {\bibfnamefont {M.~F.}\ \bibnamefont {Colombano}}, \bibinfo
  {author} {\bibfnamefont {S.}~\bibnamefont {Zanotto}}, \bibinfo {author}
  {\bibfnamefont {L.}~\bibnamefont {Vicarelli}}, \bibinfo {author}
  {\bibfnamefont {M.}~\bibnamefont {Cecchini}}, \bibinfo {author}
  {\bibfnamefont {A.}~\bibnamefont {Griol}}, \bibinfo {author} {\bibfnamefont
  {D.}~\bibnamefont {Navarro-Urrios}}, \bibinfo {author} {\bibfnamefont
  {C.}~\bibnamefont {Sotomayor-Torres}}, \bibinfo {author} {\bibfnamefont
  {A.}~\bibnamefont {Martinez}},\ and\ \bibinfo {author} {\bibfnamefont
  {J.}~\bibnamefont {Ahopelto}},\ }\bibfield  {title} {\bibinfo {title}
  {Microwave-to-optics conversion using a mechanical oscillator in its quantum
  ground state},\ }\href@noop {} {\bibfield  {journal} {\bibinfo  {journal}
  {Phys. Rev. Appl.}\ }\textbf {\bibinfo {volume} {14}},\ \bibinfo {pages}
  {014504} (\bibinfo {year} {2020})}\BibitemShut {NoStop}%
\bibitem [{\citenamefont {Forsch}\ \emph {et~al.}(2019)\citenamefont {Forsch},
  \citenamefont {Stockill}, \citenamefont {Wallucks}, \citenamefont
  {Marinković}, \citenamefont {G\"{a}rtner}, \citenamefont {Norte},
  \citenamefont {van Otten}, \citenamefont {Fiore}, \citenamefont
  {Srinivasan},\ and\ \citenamefont {Gr\"{o}blacher}}]{Forsch2020}%
  \BibitemOpen
  \bibfield  {author} {\bibinfo {author} {\bibfnamefont {M.}~\bibnamefont
  {Forsch}}, \bibinfo {author} {\bibfnamefont {R.}~\bibnamefont {Stockill}},
  \bibinfo {author} {\bibfnamefont {A.}~\bibnamefont {Wallucks}}, \bibinfo
  {author} {\bibfnamefont {I.}~\bibnamefont {Marinković}}, \bibinfo {author}
  {\bibfnamefont {C.}~\bibnamefont {G\"{a}rtner}}, \bibinfo {author}
  {\bibfnamefont {R.~A.}\ \bibnamefont {Norte}}, \bibinfo {author}
  {\bibfnamefont {F.}~\bibnamefont {van Otten}}, \bibinfo {author}
  {\bibfnamefont {A.}~\bibnamefont {Fiore}}, \bibinfo {author} {\bibfnamefont
  {K.}~\bibnamefont {Srinivasan}},\ and\ \bibinfo {author} {\bibfnamefont
  {S.}~\bibnamefont {Gr\"{o}blacher}},\ }\bibfield  {title} {\bibinfo {title}
  {Microwave-to-optics conversion using a mechanical oscillator in its quantum
  ground state},\ }\href@noop {} {\bibfield  {journal} {\bibinfo  {journal}
  {Nat. Phys.}\ }\textbf {\bibinfo {volume} {16}},\ \bibinfo {pages} {69}
  (\bibinfo {year} {2019})}\BibitemShut {NoStop}%
\bibitem [{\citenamefont {Navarro-Urrios}\ \emph {et~al.}(2022)\citenamefont
  {Navarro-Urrios}, \citenamefont {Colombano}, \citenamefont {Arregui},
  \citenamefont {Madiot}, \citenamefont {Pitanti}, \citenamefont {Griol},
  \citenamefont {Makkonen}, \citenamefont {Ahopelto}, \citenamefont
  {Sotomayor-Torres},\ and\ \citenamefont {Martínez}}]{Navarro2022}%
  \BibitemOpen
  \bibfield  {author} {\bibinfo {author} {\bibfnamefont {D.}~\bibnamefont
  {Navarro-Urrios}}, \bibinfo {author} {\bibfnamefont {M.~F.}\ \bibnamefont
  {Colombano}}, \bibinfo {author} {\bibfnamefont {G.}~\bibnamefont {Arregui}},
  \bibinfo {author} {\bibfnamefont {G.}~\bibnamefont {Madiot}}, \bibinfo
  {author} {\bibfnamefont {A.}~\bibnamefont {Pitanti}}, \bibinfo {author}
  {\bibfnamefont {A.}~\bibnamefont {Griol}}, \bibinfo {author} {\bibfnamefont
  {T.}~\bibnamefont {Makkonen}}, \bibinfo {author} {\bibfnamefont
  {J.}~\bibnamefont {Ahopelto}}, \bibinfo {author} {\bibfnamefont {C.~M.}\
  \bibnamefont {Sotomayor-Torres}},\ and\ \bibinfo {author} {\bibfnamefont
  {A.}~\bibnamefont {Martínez}},\ }\bibfield  {title} {\bibinfo {title}
  {Room-temperature silicon platform for ghz-frequency
  nanoelectro-opto-mechanical systems},\ }\href@noop {} {\bibfield  {journal}
  {\bibinfo  {journal} {ACS Phot.}\ }\textbf {\bibinfo {volume} {9}},\ \bibinfo
  {pages} {413} (\bibinfo {year} {2022})}\BibitemShut {NoStop}%
\bibitem [{\citenamefont {Khelif}\ \emph {et~al.}(2003)\citenamefont {Khelif},
  \citenamefont {Choujaa}, \citenamefont {Djafari-Roujani}, \citenamefont
  {Wilm}, \citenamefont {Ballandras},\ and\ \citenamefont
  {Laude}}]{Khelif2003}%
  \BibitemOpen
  \bibfield  {author} {\bibinfo {author} {\bibfnamefont {A.}~\bibnamefont
  {Khelif}}, \bibinfo {author} {\bibfnamefont {A.}~\bibnamefont {Choujaa}},
  \bibinfo {author} {\bibfnamefont {B.}~\bibnamefont {Djafari-Roujani}},
  \bibinfo {author} {\bibfnamefont {M.}~\bibnamefont {Wilm}}, \bibinfo {author}
  {\bibfnamefont {S.}~\bibnamefont {Ballandras}},\ and\ \bibinfo {author}
  {\bibfnamefont {V.}~\bibnamefont {Laude}},\ }\bibfield  {title} {\bibinfo
  {title} {Trapping and guiding of acoustic waves by defect modes in a
  full-band-gap ultrasonic crystal},\ }\href@noop {} {\bibfield  {journal}
  {\bibinfo  {journal} {Phys. Rev. B}\ }\textbf {\bibinfo {volume} {68}},\
  \bibinfo {pages} {214301} (\bibinfo {year} {2003})}\BibitemShut {NoStop}%
\bibitem [{\citenamefont {Fang}\ \emph {et~al.}(2016)\citenamefont {Fang},
  \citenamefont {Matheny}, \citenamefont {Luan},\ and\ \citenamefont
  {Painter}}]{Fang2016}%
  \BibitemOpen
  \bibfield  {author} {\bibinfo {author} {\bibfnamefont {K.}~\bibnamefont
  {Fang}}, \bibinfo {author} {\bibfnamefont {M.~H.}\ \bibnamefont {Matheny}},
  \bibinfo {author} {\bibfnamefont {X.}~\bibnamefont {Luan}},\ and\ \bibinfo
  {author} {\bibfnamefont {O.}~\bibnamefont {Painter}},\ }\bibfield  {title}
  {\bibinfo {title} {Optical transduction and routing of microwave phonons in
  cavity-optomechanical circuits},\ }\href@noop {} {\bibfield  {journal}
  {\bibinfo  {journal} {Nat. Phot.}\ }\textbf {\bibinfo {volume} {10}},\
  \bibinfo {pages} {489} (\bibinfo {year} {2016})}\BibitemShut {NoStop}%
\bibitem [{\citenamefont {Kokkonnen}\ and\ \citenamefont
  {Kaivola}(2007)}]{Kokkonnen2007}%
  \BibitemOpen
  \bibfield  {author} {\bibinfo {author} {\bibfnamefont {K.}~\bibnamefont
  {Kokkonnen}}\ and\ \bibinfo {author} {\bibfnamefont {M.}~\bibnamefont
  {Kaivola}},\ }\bibfield  {title} {\bibinfo {title} {Scattering of surface
  acoustic waves by a phononic crystal revealed by heterodyne interferometry},\
  }\href@noop {} {\bibfield  {journal} {\bibinfo  {journal} {Appl. Phys.
  Lett.}\ }\textbf {\bibinfo {volume} {91}},\ \bibinfo {pages} {083517}
  (\bibinfo {year} {2007})}\BibitemShut {NoStop}%
\bibitem [{\citenamefont {Profunser}\ \emph {et~al.}(2009)\citenamefont
  {Profunser}, \citenamefont {Muramoto}, \citenamefont {Matsuda}, \citenamefont
  {Wright},\ and\ \citenamefont {Lang}}]{Profunser2009}%
  \BibitemOpen
  \bibfield  {author} {\bibinfo {author} {\bibfnamefont {D.~M.}\ \bibnamefont
  {Profunser}}, \bibinfo {author} {\bibfnamefont {E.}~\bibnamefont {Muramoto}},
  \bibinfo {author} {\bibfnamefont {O.}~\bibnamefont {Matsuda}}, \bibinfo
  {author} {\bibfnamefont {O.~B.}\ \bibnamefont {Wright}},\ and\ \bibinfo
  {author} {\bibfnamefont {U.}~\bibnamefont {Lang}},\ }\bibfield  {title}
  {\bibinfo {title} {Dynamic visualization of surface acoustic waves on a
  two-dimensional phononic crystal},\ }\href@noop {} {\bibfield  {journal}
  {\bibinfo  {journal} {Phys. Rev. B}\ }\textbf {\bibinfo {volume} {80}},\
  \bibinfo {pages} {014301} (\bibinfo {year} {2009})}\BibitemShut {NoStop}%
\bibitem [{\citenamefont {Achaoui}\ \emph {et~al.}(2011)\citenamefont
  {Achaoui}, \citenamefont {Khelif}, \citenamefont {Benchabane}, \citenamefont
  {Robert},\ and\ \citenamefont {Laude}}]{Achaoui2011}%
  \BibitemOpen
  \bibfield  {author} {\bibinfo {author} {\bibfnamefont {Y.}~\bibnamefont
  {Achaoui}}, \bibinfo {author} {\bibfnamefont {A.}~\bibnamefont {Khelif}},
  \bibinfo {author} {\bibfnamefont {S.}~\bibnamefont {Benchabane}}, \bibinfo
  {author} {\bibfnamefont {L.}~\bibnamefont {Robert}},\ and\ \bibinfo {author}
  {\bibfnamefont {V.}~\bibnamefont {Laude}},\ }\bibfield  {title} {\bibinfo
  {title} {Experimental observation of locally-resonant and bragg band gaps for
  surface guided waves in a phononic crystal of pillars},\ }\href@noop {}
  {\bibfield  {journal} {\bibinfo  {journal} {Phys. Rev. B}\ }\textbf {\bibinfo
  {volume} {83}},\ \bibinfo {pages} {104201} (\bibinfo {year}
  {2011})}\BibitemShut {NoStop}%
\bibitem [{\citenamefont {Kurosu}\ \emph {et~al.}(2018)\citenamefont {Kurosu},
  \citenamefont {Hatanaka}, \citenamefont {Onomitsu},\ and\ \citenamefont
  {Yamaguchi}}]{Kurosu2018}%
  \BibitemOpen
  \bibfield  {author} {\bibinfo {author} {\bibfnamefont {M.}~\bibnamefont
  {Kurosu}}, \bibinfo {author} {\bibfnamefont {D.}~\bibnamefont {Hatanaka}},
  \bibinfo {author} {\bibfnamefont {K.}~\bibnamefont {Onomitsu}},\ and\
  \bibinfo {author} {\bibfnamefont {H.}~\bibnamefont {Yamaguchi}},\ }\bibfield
  {title} {\bibinfo {title} {On-chip temporal focusing of elastic waves in a
  phononic crystal waveguide},\ }\href@noop {} {\bibfield  {journal} {\bibinfo
  {journal} {Nat. Comm.}\ }\textbf {\bibinfo {volume} {9}},\ \bibinfo {pages}
  {1331} (\bibinfo {year} {2018})}\BibitemShut {NoStop}%
\bibitem [{\citenamefont {Rabe}\ and\ \citenamefont {Arnold}(1994)}]{Rabe1994}%
  \BibitemOpen
  \bibfield  {author} {\bibinfo {author} {\bibfnamefont {U.}~\bibnamefont
  {Rabe}}\ and\ \bibinfo {author} {\bibfnamefont {W.}~\bibnamefont {Arnold}},\
  }\bibfield  {title} {\bibinfo {title} {Acoustic microscopy by atomic force
  microscopy},\ }\href@noop {} {\bibfield  {journal} {\bibinfo  {journal}
  {Appl. Phys. Lett.}\ }\textbf {\bibinfo {volume} {64}},\ \bibinfo {pages}
  {1493} (\bibinfo {year} {1994})}\BibitemShut {NoStop}%
\bibitem [{\citenamefont {Chilla}\ \emph {et~al.}(1997)\citenamefont {Chilla},
  \citenamefont {Hesjedal},\ and\ \citenamefont {Fr\:{o}hlich}}]{Chilla1997}%
  \BibitemOpen
  \bibfield  {author} {\bibinfo {author} {\bibfnamefont {E.}~\bibnamefont
  {Chilla}}, \bibinfo {author} {\bibfnamefont {T.}~\bibnamefont {Hesjedal}},\
  and\ \bibinfo {author} {\bibfnamefont {H.~J.}\ \bibnamefont {Fr\:{o}hlich}},\
  }\bibfield  {title} {\bibinfo {title} {Nanoscale determination of phase
  velocity by scanning acoustic force microscopy},\ }\href@noop {} {\bibfield
  {journal} {\bibinfo  {journal} {Phys. Rev. B}\ }\textbf {\bibinfo {volume}
  {55}},\ \bibinfo {pages} {15858} (\bibinfo {year} {1997})}\BibitemShut
  {NoStop}%
\bibitem [{\citenamefont {Kubat}\ \emph {et~al.}(2004)\citenamefont {Kubat},
  \citenamefont {Ruile}, \citenamefont {Hesjedal}, \citenamefont {Stotz},
  \citenamefont {Rosler},\ and\ \citenamefont {Reindl}}]{Kubat2004}%
  \BibitemOpen
  \bibfield  {author} {\bibinfo {author} {\bibfnamefont {F.}~\bibnamefont
  {Kubat}}, \bibinfo {author} {\bibfnamefont {W.}~\bibnamefont {Ruile}},
  \bibinfo {author} {\bibfnamefont {T.}~\bibnamefont {Hesjedal}}, \bibinfo
  {author} {\bibfnamefont {J.}~\bibnamefont {Stotz}}, \bibinfo {author}
  {\bibfnamefont {U.}~\bibnamefont {Rosler}},\ and\ \bibinfo {author}
  {\bibfnamefont {L.~M.}\ \bibnamefont {Reindl}},\ }\bibfield  {title}
  {\bibinfo {title} {Calculation and experimental verification of the acoustic
  stress at ghz frequencies in saw resonators},\ }\href@noop {} {\bibfield
  {journal} {\bibinfo  {journal} {IEEE T. Ultrason. Ferr.}\ }\textbf {\bibinfo
  {volume} {51}},\ \bibinfo {pages} {1437} (\bibinfo {year}
  {2004})}\BibitemShut {NoStop}%
\bibitem [{\citenamefont {Hesjedal}(2010)}]{Hesjedal2010}%
  \BibitemOpen
  \bibfield  {author} {\bibinfo {author} {\bibfnamefont {T.}~\bibnamefont
  {Hesjedal}},\ }\bibfield  {title} {\bibinfo {title} {Surface acoustic
  wave-assisted scanning probe microscopy—a summary},\ }\href@noop {}
  {\bibfield  {journal} {\bibinfo  {journal} {Rep. Prog. Phys.}\ }\textbf
  {\bibinfo {volume} {73}},\ \bibinfo {pages} {016102} (\bibinfo {year}
  {2010})}\BibitemShut {NoStop}%
\bibitem [{\citenamefont {Hu}\ \emph {et~al.}(2011)\citenamefont {Hu},
  \citenamefont {Su},\ and\ \citenamefont {Arnold}}]{Hu2011}%
  \BibitemOpen
  \bibfield  {author} {\bibinfo {author} {\bibfnamefont {S.}~\bibnamefont
  {Hu}}, \bibinfo {author} {\bibfnamefont {C.}~\bibnamefont {Su}},\ and\
  \bibinfo {author} {\bibfnamefont {W.}~\bibnamefont {Arnold}},\ }\bibfield
  {title} {\bibinfo {title} {Imaging of subsurface structures using atomic
  force acoustic microscopy at ghz frequencies},\ }\href@noop {} {\bibfield
  {journal} {\bibinfo  {journal} {J. Appl. Phys.}\ }\textbf {\bibinfo {volume}
  {109}},\ \bibinfo {pages} {084324} (\bibinfo {year} {2011})}\BibitemShut
  {NoStop}%
\bibitem [{\citenamefont {Hellemann}\ \emph {et~al.}(2022)\citenamefont
  {Hellemann}, \citenamefont {M\"uller}, \citenamefont {Msall}, \citenamefont
  {Santos},\ and\ \citenamefont {Ludwig}}]{Hellemann2022}%
  \BibitemOpen
  \bibfield  {author} {\bibinfo {author} {\bibfnamefont {J.}~\bibnamefont
  {Hellemann}}, \bibinfo {author} {\bibfnamefont {F.}~\bibnamefont {M\"uller}},
  \bibinfo {author} {\bibfnamefont {M.}~\bibnamefont {Msall}}, \bibinfo
  {author} {\bibfnamefont {P.~V.}\ \bibnamefont {Santos}},\ and\ \bibinfo
  {author} {\bibfnamefont {S.}~\bibnamefont {Ludwig}},\ }\bibfield  {title}
  {\bibinfo {title} {Determining amplitudes of standing surface acoustic waves
  via atomic force microscopy},\ }\href
  {https://doi.org/10.1103/PhysRevApplied.17.044024} {\bibfield  {journal}
  {\bibinfo  {journal} {Phys. Rev. Appl.}\ }\textbf {\bibinfo {volume} {17}},\
  \bibinfo {pages} {044024} (\bibinfo {year} {2022})}\BibitemShut {NoStop}%
\bibitem [{\citenamefont {Conte}\ \emph {et~al.}(2022)\citenamefont {Conte},
  \citenamefont {Vicarelli}, \citenamefont {Zanotto},\ and\ \citenamefont
  {Pitanti}}]{Conte2022}%
  \BibitemOpen
  \bibfield  {author} {\bibinfo {author} {\bibfnamefont {G.}~\bibnamefont
  {Conte}}, \bibinfo {author} {\bibfnamefont {L.}~\bibnamefont {Vicarelli}},
  \bibinfo {author} {\bibfnamefont {S.}~\bibnamefont {Zanotto}},\ and\ \bibinfo
  {author} {\bibfnamefont {A.}~\bibnamefont {Pitanti}},\ }\bibfield  {title}
  {\bibinfo {title} {Mechanical mode engineering with orthotropic metamaterial
  membranes},\ }\href {https://doi.org/https://doi.org/10.1002/admt.202200337}
  {\bibfield  {journal} {\bibinfo  {journal} {Advanced Materials Technologies}\
  }\textbf {\bibinfo {volume} {7}},\ \bibinfo {pages} {2200337} (\bibinfo
  {year} {2022})},\ \Eprint
  {https://arxiv.org/abs/https://onlinelibrary.wiley.com/doi/pdf/10.1002/admt.202200337}
  {https://onlinelibrary.wiley.com/doi/pdf/10.1002/admt.202200337} \BibitemShut
  {NoStop}%
\bibitem [{\citenamefont {Sharahi}\ \emph {et~al.}(2021)\citenamefont
  {Sharahi}, \citenamefont {Janmaleki}, \citenamefont {Tetard}, \citenamefont
  {Kim}, \citenamefont {Sadeghian},\ and\ \citenamefont
  {Verbiest}}]{Sharahi2021}%
  \BibitemOpen
  \bibfield  {author} {\bibinfo {author} {\bibfnamefont {H.~J.}\ \bibnamefont
  {Sharahi}}, \bibinfo {author} {\bibfnamefont {M.}~\bibnamefont {Janmaleki}},
  \bibinfo {author} {\bibfnamefont {L.}~\bibnamefont {Tetard}}, \bibinfo
  {author} {\bibfnamefont {S.}~\bibnamefont {Kim}}, \bibinfo {author}
  {\bibfnamefont {H.}~\bibnamefont {Sadeghian}},\ and\ \bibinfo {author}
  {\bibfnamefont {G.~J.}\ \bibnamefont {Verbiest}},\ }\bibfield  {title}
  {\bibinfo {title} {Acoustic subsurface-atomic force microscopy:
  Three-dimensional imaging at the nanoscale},\ }\href@noop {} {\bibfield
  {journal} {\bibinfo  {journal} {J. Appl. Phys.}\ }\textbf {\bibinfo {volume}
  {129}},\ \bibinfo {pages} {030901} (\bibinfo {year} {2021})}\BibitemShut
  {NoStop}%
\bibitem [{\citenamefont {de~Lima}\ and\ \citenamefont
  {Santos}(2005)}]{deLima2005_rev}%
  \BibitemOpen
  \bibfield  {author} {\bibinfo {author} {\bibfnamefont {M.~M.}\ \bibnamefont
  {de~Lima}}\ and\ \bibinfo {author} {\bibfnamefont {P.~V.}\ \bibnamefont
  {Santos}},\ }\bibfield  {title} {\bibinfo {title} {Modulation of photonic
  structures by surface acoustic waves},\ }\href@noop {} {\bibfield  {journal}
  {\bibinfo  {journal} {Rep. Prog. Phys.}\ }\textbf {\bibinfo {volume} {68}},\
  \bibinfo {pages} {1639} (\bibinfo {year} {2005})}\BibitemShut {NoStop}%
\bibitem [{\citenamefont {Behme}\ and\ \citenamefont
  {Hesjedal}(2001)}]{Behme2001}%
  \BibitemOpen
  \bibfield  {author} {\bibinfo {author} {\bibfnamefont {G.}~\bibnamefont
  {Behme}}\ and\ \bibinfo {author} {\bibfnamefont {T.}~\bibnamefont
  {Hesjedal}},\ }\bibfield  {title} {\bibinfo {title} {Influence of surface
  acoustic waves on lateral forces in scanning force microscopies},\
  }\href@noop {} {\bibfield  {journal} {\bibinfo  {journal} {J. Appl. Phys.}\
  }\textbf {\bibinfo {volume} {89}},\ \bibinfo {pages} {4850} (\bibinfo {year}
  {2001})}\BibitemShut {NoStop}%
\end{thebibliography}%

\end{document}